\documentclass[twocolumn,showpacs,prd,floatfix,axodraw]{revtex4}
\usepackage{mathrsfs}
\usepackage{graphicx,,booktabs,bm}
\usepackage{overpic}
\usepackage{color}
\usepackage{amssymb}
\usepackage{hyperref}
\usepackage{soul}

\usepackage{mathrsfs,bm,amsmath,amssymb}
\usepackage{longtable,lscape}
\usepackage{txfonts}
\usepackage{amssymb}
\usepackage{indentfirst}
\usepackage{graphicx,,booktabs}
\usepackage{multirow,ulem}

\usepackage{color}
\usepackage{amssymb}

\begin{document}
\title{Strong decays of the explicitly exotic doubly charmed $DDK$ bound state }

\author{Yin Huang}
\affiliation{School of Physics, Beihang University, Beijing 100191, China}

\author{Ming-Zhu Liu}
\affiliation{School of Physics, Beihang University, Beijing 100191, China}

\author{Ya-Wen Pan}
\affiliation{School of Physics, Beihang University, Beijing 100191, China}

\author{Li-Sheng Geng}
\email{lisheng.geng@buaa.edu.cn}
\affiliation{School of Physics \&
Beijing Advanced Innovation Center for Big Data-based Precision Medicine, Beihang University, Beijing100191, China.}
\affiliation{School of Physics, Zhengzhou University, Zhengzhou, Henan 450001, China}

\author{A. Mart\'inez Torres}
\email{amartine@if.usp.br}
\affiliation{Instituto de F\'isica, Universidade de S\~ao Paulo, C.P. 66318, 05389-970 S\~ao
Paulo, S\~ao Paulo, Brazil.}
\affiliation{School of Physics and Nuclear Energy Engineering \& Beijing Key Laboratory of Advanced Nuclear Materials and Physics, Beihang University, Beijing 100191, China}

\author{K. P. Khemchandani}
\email{kanchan.khemchandani@unifesp.br}
\affiliation{Universidade Federal de S\~ao Paulo, C.P. 01302-907, S\~ao Paulo, Brazil.}
\affiliation{School of Physics and Nuclear Energy Engineering \& Beijing Key Laboratory of Advanced Nuclear Materials and Physics, Beihang University, Beijing 100191, China}

\begin{abstract}
Nowadays, it is generally accepted that the $DK$ interaction in isospin zero is strongly attractive and the $D_{s0}^*(2317)$ can be described as a $DK$ molecular state. Recent studies show that  the three-body $DDK$ system binds as well with a binding energy about 60$\sim$70 MeV.  The $DDK$ bound state has isospin $1/2$ and spin-parity $0^-$. If discovered either experimentally or in lattice QCD, it will not only provide further support on the molecular nature of the $D_{s0}^*(2317)$, but
also provide a way to understand other exotic hadrons expected to be of molecular nature. In the present work, we study its two-body strong decay widths via triangle diagrams. We find that the partial decay width into $DD_s\pi$ is at the order of $2\sim3$ MeV, which seems to be within the reach of the current experiments such as BelleII. As a result, we strongly recommend this decay channel of the $DDK$ bound state to be searched for experimentally.
\end{abstract}

\date{\today}


\maketitle
\section{Introduction}
In 2003, the BaBar Collaboration observed a narrow state in the inclusive $D^+_s\pi^0$  invariant mass distribution
of the $e^+e^-$ collision at energies near 10.6 GeV~\cite{Aubert:2003fg}, i.e., the $D_{s0}^*(2317)$ ($D_{s0}$ for short in the present work), which was later confirmed
by the CLEO Collaboration~\cite{Besson:2003cp} and the Belle Collaboration~\cite{Krokovny:2003zq}. Because the low mass, small width, and decay mode of the $D_{s0}$ are  quite different from those of
a conventional $J^P=0^+$ $c\bar{s}$ state in the naive quark model, its nature has remained a topic of tremendous theoretical interests
ever since its discovery~\cite{Bardeen:2003kt,Nowak:2003ra,vanBeveren:2003kd,Dai:2003yg,Narison:2003td,Szczepaniak:2003vy,Browder:2003fk,Barnes:2003dj,Cheng:2003kg,Chen:2004dy,Dmitrasinovic:2005gc,Zhang:2018mnm,Terasaki:2003qa,Maiani:2004vq,Kolomeitsev:2003ac,Hofmann:2003je,Guo:2006fu,Gamermann:2006nm,Guo:2008gp,Guo:2009ct,Cleven:2010aw,MartinezTorres:2011pr,Yao:2015qia,Guo:2015dha,Albaladejo:2016lbb,Du:2017ttu,Guo:2018kno,Albaladejo:2018mhb,Altenbuchinger:2013gaa,Altenbuchinger:2013vwa,Geng:2010vw,Wang:2012bu,Liu:2009uz,Guo:2018tjx}. In recent years, the importance  of the $DK$ interaction in forming the $D_{s0}$ has been  confirmed by  lattice QCD simulations~\cite{Liu:2012zya,Mohler:2013rwa, Lang:2014yfa,Torres:2014vna,Bali:2017pdv}. See Ref.~\cite{Guo:2019dpg} for a short summary of
the theoretical, experimental, and lattice QCD supports for the molecular interpretation of the $D_{s0}$ as a $DK$ bound state.

If the $D_{s0}$ is indeed (dominantly) a $DK$ bound state, a natural question to ask is whether the $DDK$  three-body system is still bound.
In Ref.~\cite{SanchezSanchez:2017xtl},  by describing the $D_{s0} D$ interaction via one kaon exchange (OKE), it was shown that
the OKE interaction is strong enough to form a $D_{s0} D$ molecular state with a binding energy of $50\sim60$ MeV, regardless whether the $D_{s0} DK$ coupling is determined by treating the $D_{s0}$ as
a $c\bar{s}$ state or a $DK$ molecule. In Ref.~\cite{MartinezTorres:2018zbl}, a study was done by explicitly considering the three-body $D(DK-D_s\pi-D_s\eta)$ system and by solving the Faddeev equations
using the two-body inputs provided by the unitarized chiral perturbation theory and the local hidden symmetry approach.
A three-body bound state was found in this latter work, with a total mass around 4140 MeV, which is an isospin doublet containing two states $(R^{++}, R^+$).  In a more recent work~\cite{Wu:2019vsy},
using the Gaussian expansion method, the existence of this state has been further confirmed  though with a lightly smaller binding energy of $\sim60-70$ MeV)and
it has been found that even the $DDDK$ or $DDD_{s0}$ system is  bound. It is interesting to note that the $D\bar{D}^*K$~\cite{Ma:2017ery,Ren:2018pcd,Ren:2019umd}, the $DKK$ and $DK\bar{K}$~\cite{Debastiani:2017vhv}, as well as the $DK\bar{N}$~\cite{Yamagata-Sekihara:2018gah} systems bind as well,  because of
the strong attraction between $D$ and $K$.

As pointed out in Ref.~\cite{MartinezTorres:2018zbl}, the three-body $DDK$ bound state  can decay strongly via diagrams such as those shown in
Fig.~\ref{mku}. In the present work, we calculate explicitly the partial decay widths from such processes, aiming to provide further motivation for the experimental search for this state.
The present manuscript is organized as follows. The theoretical
formalism is explained in Sec. II. The predicted partial
decay widths are presented in Sec. III, followed by a short summary in Sec. IV.

 \section{Theoretical formlism}
\begin{figure}[htbp]
\begin{center}
\includegraphics[scale=0.45]{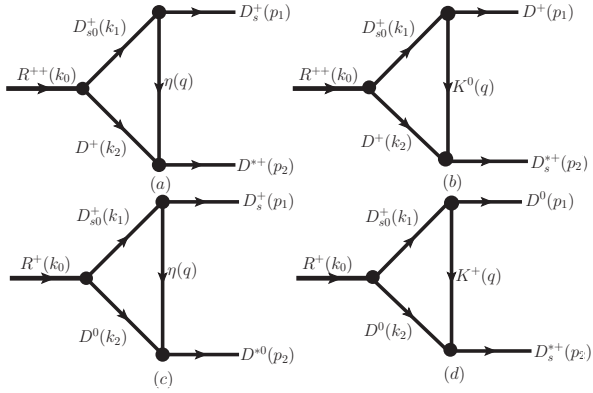}
\caption{Tirangle diagrams representing the decay of the $R^{++}$ state to $D_s^{+}D^{*+}$ and $D^{+}D_s^{*+}$(a-b), and $R^{+}$ state to $D_s^{+}D^{*0}$ and $D^{0}D_s^{*+}$(c-d).}\label{mku}
\end{center}
\end{figure}

In the following, we focus on the doubly charged state $R^{++}$. Due to isospin symmetry, the decay width of its isospin partner $R^+$
can be calculated analogously and only small differences are expected because of the slightly different masses of its molecular components. As mentioned in Ref.~\cite{MartinezTorres:2018zbl}, though the $R^{++}$ is a bound state of the $DDK$ system or $D_{s0}$D system, it is possible for such a state to decay strongly.  Keeping in mind that the $D_{s0}^{+}$  is observed in the
inclusive $D^+_s\pi^0$ invariant mass distribution, which violates isospin, the $R^{++}$ can decay via $R^{++}\equiv D_{s0}^{+} D^+\rightarrow (D^+_s\pi^0) D^+$.
An alternative  process, without involving isospin breaking, is via triangle diagrams such as those shown in Fig.~\ref{mku}. These processes conserve isospin and therefore should be the dominant ones, as compared to the ones that
violate isospin. In the following, we explain how to calculate the four diagrams shown in Fig.~\ref{mku}.

In order to calculate the Feynman diagrams shown in Fig.~\ref{mku}, we need to determine the relevant vertices. For the vertex of $R^{++} D_{s0}^{+}D^{+}$, since the $R^{++}$ can be treated as
a bound state of $D_{s0}^{+} D^{+}$~\cite{SanchezSanchez:2017xtl}, this coupling can be determined by the Weinberg compositeness condition. In the
present work, we adopt the method developed in Refs.~\cite{Faessler:2007gv,Faessler:2007us,Dong:2008gb,Dong:2009uf,Dong:2009yp,Dong:2017rmg,Dong:2014ksa,Dong:2014zka,Dong:2013kta,Dong:2013iqa,Dong:2013rsa,Dong:2012hc,Dong:2011ys,Dong:2010xv,Dong:2010gu,Dong:2009tg,Dong:2017gaw}.  In this framework, the interacting Lagrangian between $R$, $D_{s0}$, and $D$ can be written as~\cite{Faessler:2007gv,Faessler:2007us}
\begin{figure}[htbp]
\begin{center}
\includegraphics[scale=0.45]{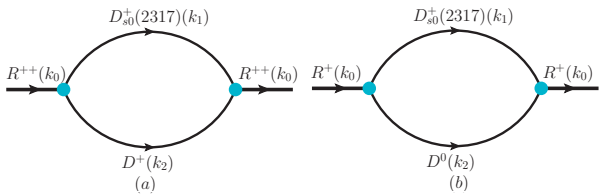}
\caption{Self-energy of the $R^{++}$ and $R^{+}$ states.}\label{mku1}
\end{center}
\end{figure}

\begin{align}
{\cal{L}}_{R}(x)=g_{RD_{s0}D}(x)R^{T}(x)&\int{}dy\Phi_{R}(y^2)D_{s0}(x+\omega_{D}y)\nonumber\\
                                        &\times{}D(x-\omega_{D_{s0}}y)+H.c.,\label{eq1}
\end{align}
where $\omega_{i}=m_{i}/(m_i+m_j)$ is  a kinematical parameter with $m_i$ and $m_j$ being the masses of the involved mesons.
In the Lagrangian of Eq.~(\ref{eq1}),  an effective correlation function $\Phi(y^2)$ is introduced to reflect the distribution of the two constituents,
$D^{+}_{s0}(2317)$ and $D^{+}(D^0)$, in the hadronic molecular $R^{++}(R^+)$ state.  The introduced correlation function also serves the purpose of making the
Feynman diagrams ultraviolate finite. Here we choose the Fourier transformation of the
correlation function in terms of a Gaussian form,
\begin{align}
\Phi(p^2)\doteq\exp(-p_E^2/\Lambda^2)
\end{align}
where $\Lambda\sim1.0$ GeV~\cite{Faessler:2007gv,Faessler:2007us,Dong:2008gb,Dong:2009uf,Dong:2009yp,Dong:2017rmg,Dong:2014ksa,Dong:2014zka,Dong:2013kta,Dong:2013iqa,Dong:2013rsa,Dong:2012hc,Dong:2011ys,Dong:2010xv,Dong:2010gu,Dong:2009tg,Dong:2017gaw} is a size parameter, which characterizes the distribution of the molecular components inside the molecule, and  $p_E$ is the Euclidean
Jacobi momentum~\cite{Faessler:2007gv,Faessler:2007us,Dong:2008gb,Dong:2009uf,Dong:2009yp,Dong:2017rmg,Dong:2014ksa,Dong:2014zka,Dong:2013kta,Dong:2013iqa,Dong:2013rsa,Dong:2012hc,Dong:2011ys,Dong:2010xv,Dong:2010gu,Dong:2009tg,Dong:2017gaw}. In the present work, we take $\Lambda=1$ GeV, unless otherwise stated.

The coupling constant $g_{RD_{s0}D}$ in Eq.~(\ref{eq1}) could be determined by the compositeness condition~\cite{Faessler:2007gv,Faessler:2007us},
where the renormalization constant of the composite particle should be zero, i.e.,
\begin{align}
Z_{R}\equiv{}1-\Sigma_{R}(m^2_{R})=0,
\end{align}
with $\Sigma_{R}(m^2_{R})$ being the derivative of the mass operator of the $R$.    The concrete form of the mass
operator of the $DDK$ bound state $R$ corresponding to the diagram in Fig.~\ref{mku1} is
\begin{align}
{\cal{W}}_{R}(k_0)&=\frac{g^2_{RD_{s0}D}}{16\pi^2}\int_0^{\infty}d\alpha\int_0^{\infty}d\beta\frac{1}{z^2}\exp\{-\frac{1}{\Lambda^2}\nonumber\\
                    &\times[-2k^2_0\omega^2_{D_{s0}}+\alpha{}m^2_{D_{s0}}+\beta(-k^2_0+m^2_D)+\frac{\Delta^2_{M}}{4z}]\}\label{mkqy},
\end{align}
where $z=2+\alpha+\beta$, $\Delta_M=-4\omega_{D_{s0}}k_0-2\beta{}k_0$, and $k_0^2=m^2_{R}$ with $k_0$, $m_{R}$ denoting the four-momenta and mass of the $R$, respectively.  Here, we set $m_{R}=m_{D_{s0}}+m_{D}-E_b$ with $E_b$ being the binding energy of $R$,  $k_1$, and $m_{D_{s0}}$ are  the four-momenta and
mass of the $D_{s0}$, and $m_{D}$ is the  mass of the $D$-meson, respectively.

In the present work, we calculate the two-body decay width of the $R$ via the triangle diagrams shown in Fig.~\ref{mku}.
To evaluate the diagrams, in addition to the Lagrangian of Eq.~(\ref{eq1}), the following effective Lagrangian terms, responsible for the interactions between
heavy-light pseudoscalar and vector mesons\textcolor{blue}{,} are needed as well~\cite{Altenbuchinger:2013vwa}
\begin{align}
{\cal{L}}_{P\phi P^{*}}=ig\langle P^{*}_{\mu}u^{\mu}P^{\dagger}-Pu^{\mu}P_{\mu}^{*\dagger}\rangle\label{eq5},
\end{align}
where $P=(D^{0},D^{+},D^{+}_s)$ and  $P^{*}=(D^{*0},D^{*+},D^{*+}_s)$, $u^{\mu}$ is the axial vector combination of the pseudoscalar-meson
fields and their derivatives,
\begin{align}
u^{\mu}=i(u^{\dagger}\partial^{\mu}u-u\partial^{\mu}u^{\dagger}),
\end{align}
where $u^2 = U = \exp(i\frac{\phi}{f_0})$, $f_0$=92.4 MeV, and the pseudoscalar- meson octet $\phi$ is represented by the $3\times3$ matrix
\begin{equation}
\phi=\sqrt{2}
\left(
  \begin{array}{ccc}
    \frac{\pi^{0}}{\sqrt{2}}+\frac{\eta}{\sqrt{6}} &    \pi^{+}                                        &       K^{+}\\
    \pi^{-}                                       &    -\frac{\pi^{0}}{\sqrt{2}}+\frac{\eta}{\sqrt{6}} &       K^{0}\\
    K^{-}                                         &    \bar{K}^{0}                                     &       -\frac{2}{\sqrt{6}}\eta
  \end{array}
\right)\label{eq7}.
\end{equation}

From Eqs.~(\ref{eq5})-(\ref{eq7}), one can easily obtain the interaction vertices $\eta{}DD^{*}$,$KDD^{*}_{s}$, and $\pi D^{*}D$.
The coupling constant $g$ can be determined from the strong decay width $\Gamma(D^{*+}\to{}D^{0}\pi^{+})=56.46\pm1.22$ keV\textcolor{blue}{,} together with
the branching ratio $BR(D^{*+}\to{}D^{0}\pi^{+})=(67.7\pm0.5)\%$~\cite{Tanabashi:2018oca}.  With the help of Eq.~(\ref{eq5}), the two
body decay width $\Gamma(D^{*+}\to\pi^{+}D^{0})$ is related to $g$ via
\begin{align}
\Gamma(D^{*+}\to\pi^{+}D^{0})=\frac{1}{12\pi}\frac{g^2}{f_0^2}\frac{|\vec{p}_{\pi}|^3}{M^2},
\end{align}
where $\vec{p}_{\pi}$ is the three-momentum of $\pi^{+}$ in the rest frame of the decaying vector meson $D^{*+}$.
Using the corresponding experimental strong decay width and the masses of the relevant particles given in 
Table~\ref{table-mass}~\cite{Tanabashi:2018oca}, we obtain $g=1.097\pm0.012$ GeV.

\begin{table}[htpb]
\centering
\caption{Masses of the relevant particles  in the present work (in units of MeV)~\cite{Tanabashi:2018oca}.}\label{table-mass}
\begin{tabular}{ccccccc}
\hline\hline
$~~~~~D^{+}$~& $~~~~~D^{0}$         & ~~~~~$\eta$          & ~~~~~$D^{*0}$     &~~~~~$D^{*\pm}$              \\ \hline
 ~~~~~1869.65  &~~~~~1864.83        & ~~~~~547.862          & ~~~~~2006.85      &~~~~~2010.26                 \\  \hline \hline
$~~~~~K^{0}$   &~~~~~ $D_s^{*\pm}$  & ~~~~~$D_{s0}^{*\pm}$ & ~~~~~$D_s^{\pm}$  &~~~~~$K^{\pm}$                \\ \hline
 ~~~~~497.611   &~~~~~ 2112.2        & ~~~~~2317.0          & ~~~~~1968.34      &~~~~~493.677                     \\  \hline \hline
\end{tabular}
\end{table}

In the chiral unitary approaches~\cite{Gamermann:2006nm,Gamermann:2007fi,Altenbuchinger:2013vwa,Guo:2006fu},  the $D_{s0}$ is found to be dynamically generated from the $DK$ and $D_{s}\eta$ $S$-wave
interactions.   As a result, the vertices $D_{s0}DK$ and $D_{s0}\eta{}D_{s}$ can be easily written as
\begin{align}
{\cal{L}}_{D_{s0}DK}&=g_{D_{s0}DK}D_{s0}DK,\\
{\cal{L}}_{D_{s0}D_{s}\eta}&=g_{D_{s0}D_{s}\eta}D_{s0}D_s\eta,
\end{align}
where the coupling of the $D_{s0}$ to $DK$ and $D_{s}\eta$ states, $g_{D_{s0}DK}$ and $g_{D_{s0}D_{s}\eta}$,
can be obtained from the coupling constant of the $D_{s0}$ to the $DK$ and $\eta{}D_{s}$ channels in isospin zero, which are found to be $g_{D_{s0}DK}=10.21$ GeV(10.203 GeV)
and $g_{D_{s0}D_{s}\eta}=6.40$ GeV(5.876 GeV) in Ref.~\cite{Gamermann:2006nm}(~\cite{Guo:2006fu}),  multiplied by the appropriate Clebsch-Gordan (CG) coefficients, namely, $g_{D^{+}_{s0}D^{+}K^{0}}=g_{D^{+}_{s0}D^{0}K^{+}}=-g_{D_{s0}DK}/\sqrt{2}$ and $g_{D^{+}_{s0}D_{s}^{+}\eta}=g_{D_{s0}D_{s}\eta}$.

With the above vertices, the amplitudes of the triangle diagrams of Fig.~\ref{mku}, evaluated in the center of mass frame of final states, are
\begin{align}
-i{\cal{M}}^{a}_{\eta}&=g_{R^{++}D^{+}_{s0}D^{+}}g_{D^{+}_{s0}D_{s}^{+}\eta}\frac{-g}{\sqrt{3}f_0}\int{}\frac{d^4q}{(2\pi)^4}\Phi[(k_1\omega_{D^{+}}\nonumber\\
                    &-k_2\omega_{D^{+}_{s0}})^2]\epsilon_{\mu}(p_2)q^{\mu}\frac{1}{q^2-m^2_{\eta}}\frac{1}{k_1^2-m^2_{D^{+}_{s0}}}\frac{1}{k_2^2-m^2_{D^{+}}},\\
-i{\cal{M}}^{b}_{K^{0}}&=g_{R^{++}D^{+}_{s0}D^{+}}g_{D^{+}_{s0}D^{+}K^{0}}\frac{-\sqrt{2}g}{f_0}\int{}\frac{d^4q}{(2\pi)^4}\Phi[(k_1\omega_D^{+}-\nonumber\\
                     &k_2\omega_{D^{+}_{s0}})^2]\epsilon_{\mu}(p_2)q^{\mu}\frac{1}{q^2-m^2_{K^{0}}}\frac{1}{k_1^2-m^2_{D^{+}_{s0}}}\frac{1}{k_2^2-m^2_{D^{+}}},\\
-i{\cal{M}}^{c}_{\eta}&=g_{R^{++}D^{+}_{s0}D^{+}}g_{D^{+}_{s0}D_{s}^{+}\eta}\frac{-g}{\sqrt{3}f_0}\int{}\frac{d^4q}{(2\pi)^4}\Phi[(k_1\omega_{D^{0}}\nonumber\\
                    &-k_2\omega_{D^{+}_{s0}})^2]\epsilon_{\mu}(p_2)q^{\mu}\frac{1}{q^2-m^2_{\eta}}\frac{1}{k_1^2-m^2_{D^{+}_{s0}}}\frac{1}{k_2^2-m^2_{D^{0}}},\\
-i{\cal{M}}^{d}_{K^{+}}&=g_{R^{++}D^{+}_{s0}D^{+}}g_{D^{+}_{s0}D^{0}K^{+}}\frac{-\sqrt{2}g}{f_0}\int{}\frac{d^4q}{(2\pi)^4}\Phi[(k_1\omega_D^{+}-\nonumber\\
                     &k_2\omega_{D^{+}_{s0}})^2]\epsilon_{\mu}(p_2)q^{\mu}\frac{1}{q^2-m^2_{K^{+}}}\frac{1}{k_1^2-m^2_{D^{+}_{s0}}}\frac{1}{k_2^2-m^2_{D^{0}}}.
\end{align}
The corresponding partial decay width then reads
\begin{equation}
d\Gamma[R\to]=\frac{1}{2J+1}\frac{1}{32\pi^2}\frac{|\vec{p}_1|}{m^2_R}\overline{|{\cal{M}}|^2}d\Omega,
\end{equation}
where $J=0$ is the total angular momentum of the initial $R$
state, the overline indicates the sum over the polarization
vectors of final hadrons, and $|\vec{p}_1|$ is the 3-momenta of
the decay products in the  rest frame of the  $(R^{++},R^{+})$ states.  Then the total decay widths of the $(R^{++},R^+)$ states are
\begin{align}
\Gamma_{R^{++}}&=\Gamma[R^{++}\to{}D_s^{+}D^{*+}]+\Gamma[R^{++}\to{}D^{+}D^{*+}_{s}],\\
      \Gamma_{R^+}   &=\Gamma[R^{+}\to{}D_s^{+}D^{*0}]+\Gamma[R^{+}\to{}D^{0}D^{*+}_{s}].
\end{align}

\section{RESULTS and Discussions}
To estimate the partial decay widths of the $R$, we first need to determine the coupling constants related to the
molecular state and its components.

\begin{figure}[htbp!]
\begin{center}
\includegraphics[scale=0.30]{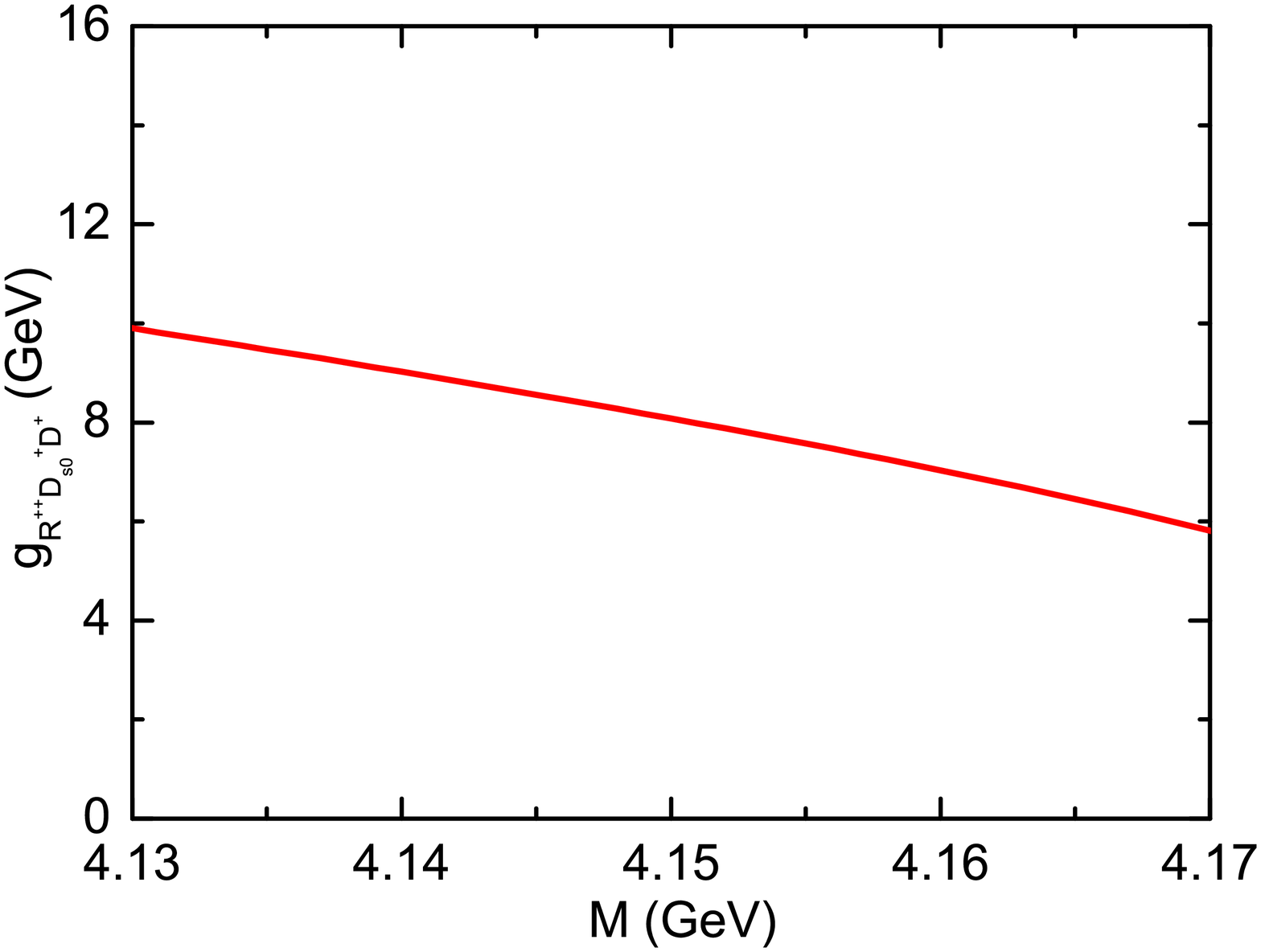}
\caption{The coupling constant $g_{R^{++} D^+_{s0} D^+}$ as a function of the mass of   $R^{++}$.}\label{couplin-constant}
\end{center}
\end{figure}

In Refs.~\cite{SanchezSanchez:2017xtl,MartinezTorres:2018zbl,Wu:2019vsy}, the $R^{++}$ state is found to have a binding energy about $15\sim45$ MeV, with respect to the $D_{s0}^{+} D^{+}$ threshold.
In this mass range, the coupling constant is dependent on the mass of the bound state $R$ as shown in Fig.~\ref{couplin-constant}. One finds that the coupling constant $g_{R^{++}D^{+}D_{s0}^{+}}$ decreases with $m_{R^{++}}$.  With a value of the mass $m_{R^{++}}=4140$ MeV, the corresponding coupling constants is $g_{R^{++}D^{+}D_{s0}^{+}}=9.02$ GeV.

\begin{figure}[htbp!]
\begin{center}
\includegraphics[scale=0.32]{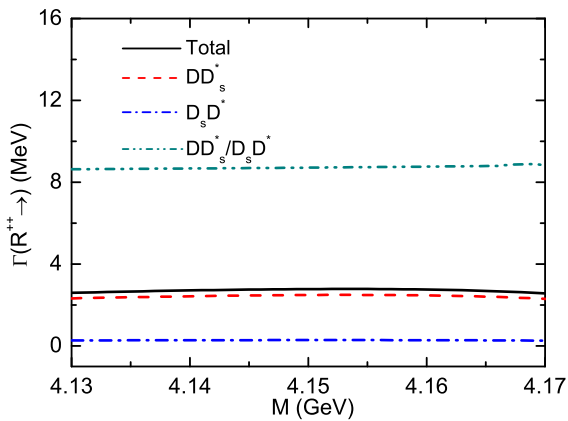}
\caption{ Total decay width (black solid line), partial decay widths of the $R^{++}\to{}D^{+}D_s^{*+}$ (red dashed line), $R^{++}\to{}D^{+}_sD^{*+}$ (blue dash-dot-dotted line), and the ratio of the partial decay widths (dark cyan dash dotted line) as a function of the mass of the $R^{++}$.}\label{mku2}
\end{center}
\end{figure}

We show the dependence of the total decay width on the  masses of the bound state $R^{++}$ in Fig.~\ref{mku2}.  One can see
that  the total decay width increases slightly with
the mass of the bound state $R^{++}$ from 4.13 to 4.17 GeV.  The predicted total decay width is small and found to be $\Gamma_{R^{++}}=2.5-2.6$  MeV.

{\color{blue}{\begin{table*}[htpb!]
\centering
\caption{ Partial decay widths of $R^{++}\to{}D^{+}_sD^{*+}$ and $D^{+}D_{s}^{*+}$ for different  $R^{++}$ masses (in units of MeV).}\label{table0}
\begin{tabular}{ccccccc}
\hline\hline
   Mode                    &~~~~~$\Lambda$(GeV)  &~~~~~  4130                   &~~~~~  4140                    &~~~~~ 4150              &~~~~~ 4160               &~~~~~4170 \\ \hline
$R^{++}\to{}D^{+}_sD^{*+}$ &~~~~~1.0             &~~~~~ $ 0.27$                 &~~~~~ $0.28$                   &~~~~~ $0.29$            &~~~~~ $0.29$             &~~~~~$0.26$   \\
$R^{++}\to{}D^{+}D_{s}^{*+}$ &~~~~~1.0           &~~~~~ $ 2.33$                 &~~~~~ $2.43$                   &~~~~~ $2.49$            &~~~~~ $2.47$             &~~~~~$2.30$   \\
 \hline               \hline
\end{tabular}
\end{table*} } }
In Fig.~\ref{mku2}, we also show the partial decay widths of the $R^{++}\to{}D^{+}D^{*+}_s$ and $D^{+}_sD^{*+}$ as a function of the
mass of the bound state $R$.  The corresponding partial decay widths for several masses of the bound state are  listed in Table.\ref{table0}.
We note that the transition $R^{++}\to{}D^{+}D_s^{*+}$ is the main decay channel, almost saturating the total width.
The corresponding partial decay widths are  $\Gamma_{R^{+}\to{}D^{+}D_s^{*+}}=2.30-2.50$ MeV
and $\Gamma_{R^{++}\to{}D^{+}_sD^{*+}}=0.26-0.29$ MeV, which yields a total decay width of $2.6\sim2.8$ MeV.  We note that the results depend only moderately on the cutoff. For instance, varying the cutoff from 0.5 to 1.5 GeV, the total decay width changes from 1.6 to 3.4 MeV within the $R^{++}$ mass range of 4.13 to 4.17 GeV.

We find that the contribution from the $\eta$ meson exchange is very small, because the $\eta D^{0}D^{*0}$ vertex,
which involves the creation or annihilation of an additional $s\bar{s}$ quark pair, is strongly suppressed.    Moreover, the main component of the $D_{s0}$(2317) is $DK$~\cite{Guo:2006fu,Gamermann:2006nm,Altenbuchinger:2013vwa,Gamermann:2007fi} and the coupling constant related to this vertex is larger than the others.   These two factors make
the contribution from the $K$ meson exchange the most important one.

In Fig.~\ref{mku2}, we also show the  ratio of the partial decay widths into $D^{+}D_s^{*+}$ and $D^{+}_sD^{*+}$.  The ratio of branching fractions is found to be of the order of $8\sim9$, and
 is almost independent of the mass of the bound state $R$.

\section{summary}
In this work, inspired by the recent series of studies that showed the likely existence of  a $DDK$ bound state, we have studied
its partial decay widths into $D_sD^{*}$ and $D D^{*}_s$. Such a decay involves the treatment of the $R$ state as a quasi-bound state of
$D_{s0}^*(2317) D$ and utilizing the Weinberg compositeness condition to determine the corresponding coupling. Our studies find a relative small total decay width, of
 the order of  $2\sim3$ MeV, mainly to $DD_s^*$, and the results depends only moderately on the  single  parameter of the method, the cutoff $\Lambda$.

The predicted decay width seems to suggest that it is possible to observe such a state at Belle or BelleII, e.g., via the inclusive  invariant mass distribution $D^+ D_s^+\pi^0$, which is quite similar to the experimental discovery of the $D_{s0}^*(2317)$ by  BaBar, Belle, and CLEO. On the other hand, its production yields at these experimental setups remain to be studied.

Recent lattice QCD studies of compact tetraquark states, see, e.g., Refs.~\cite{Francis:2016hui,Leskovec:2019ioa}, suggest that a study of the $DDK$ bound state in
terms of its minimal quark content $cc\bar{q}\bar{s}$
 might be within the reach of the state of the art of lattice QCD simulations, even taking explicitly into account its three-meson molecular nature~\cite{Horz:2019rrn}.

\section*{Acknowledgements}

LSG thanks Chen-Ping Shen and Manuel Pavon Valderrama for some stimulating discussions. This work was partly supported the National Natural Science Foundation of China (NSFC) under Grants Nos. 11975041, 11735003, and Conselho Nacional de Desenvolvimento Cient\'ifico e Tecnolo\'ogico (CNPq) under Grant Nos. 310759/2016-1 and 311524/2016-8.


\bibliography{DDDK}

\begin{thebibliography}{73}
\expandafter\ifx\csname natexlab\endcsname\relax\def\natexlab#1{#1}\fi
\expandafter\ifx\csname bibnamefont\endcsname\relax
  \def\bibnamefont#1{#1}\fi
\expandafter\ifx\csname bibfnamefont\endcsname\relax
  \def\bibfnamefont#1{#1}\fi
\expandafter\ifx\csname citenamefont\endcsname\relax
  \def\citenamefont#1{#1}\fi
\expandafter\ifx\csname url\endcsname\relax
  \def\url#1{\texttt{#1}}\fi
\expandafter\ifx\csname urlprefix\endcsname\relax\def\urlprefix{URL }\fi
\providecommand{\bibinfo}[2]{#2}
\providecommand{\eprint}[2][]{\url{#2}}

\bibitem[{\citenamefont{Aubert et~al.}(2003)}]{Aubert:2003fg}
\bibinfo{author}{\bibfnamefont{B.}~\bibnamefont{Aubert}} \bibnamefont{et~al.}
  (\bibinfo{collaboration}{BaBar}), \bibinfo{journal}{Phys. Rev. Lett.}
  \textbf{\bibinfo{volume}{90}}, \bibinfo{pages}{242001}
  (\bibinfo{year}{2003}), \eprint{hep-ex/0304021}.

\bibitem[{\citenamefont{Besson et~al.}(2003)}]{Besson:2003cp}
\bibinfo{author}{\bibfnamefont{D.}~\bibnamefont{Besson}} \bibnamefont{et~al.}
  (\bibinfo{collaboration}{CLEO}), \bibinfo{journal}{Phys. Rev.}
  \textbf{\bibinfo{volume}{D68}}, \bibinfo{pages}{032002}
  (\bibinfo{year}{2003}), \bibinfo{note}{[Erratum: Phys.
  Rev.D75,119908(2007)]}, \eprint{hep-ex/0305100}.

\bibitem[{\citenamefont{Krokovny et~al.}(2003)}]{Krokovny:2003zq}
\bibinfo{author}{\bibfnamefont{P.}~\bibnamefont{Krokovny}} \bibnamefont{et~al.}
  (\bibinfo{collaboration}{Belle}), \bibinfo{journal}{Phys. Rev. Lett.}
  \textbf{\bibinfo{volume}{91}}, \bibinfo{pages}{262002}
  (\bibinfo{year}{2003}), \eprint{hep-ex/0308019}.

\bibitem[{\citenamefont{Bardeen et~al.}(2003)\citenamefont{Bardeen, Eichten,
  and Hill}}]{Bardeen:2003kt}
\bibinfo{author}{\bibfnamefont{W.~A.} \bibnamefont{Bardeen}},
  \bibinfo{author}{\bibfnamefont{E.~J.} \bibnamefont{Eichten}},
  \bibnamefont{and} \bibinfo{author}{\bibfnamefont{C.~T.} \bibnamefont{Hill}},
  \bibinfo{journal}{Phys. Rev.} \textbf{\bibinfo{volume}{D68}},
  \bibinfo{pages}{054024} (\bibinfo{year}{2003}), \eprint{hep-ph/0305049}.

\bibitem[{\citenamefont{Nowak et~al.}(2004)\citenamefont{Nowak, Rho, and
  Zahed}}]{Nowak:2003ra}
\bibinfo{author}{\bibfnamefont{M.~A.} \bibnamefont{Nowak}},
  \bibinfo{author}{\bibfnamefont{M.}~\bibnamefont{Rho}}, \bibnamefont{and}
  \bibinfo{author}{\bibfnamefont{I.}~\bibnamefont{Zahed}},
  \bibinfo{journal}{Acta Phys. Polon.} \textbf{\bibinfo{volume}{B35}},
  \bibinfo{pages}{2377} (\bibinfo{year}{2004}), \eprint{hep-ph/0307102}.

\bibitem[{\citenamefont{van Beveren and Rupp}(2003)}]{vanBeveren:2003kd}
\bibinfo{author}{\bibfnamefont{E.}~\bibnamefont{van Beveren}} \bibnamefont{and}
  \bibinfo{author}{\bibfnamefont{G.}~\bibnamefont{Rupp}},
  \bibinfo{journal}{Phys. Rev. Lett.} \textbf{\bibinfo{volume}{91}},
  \bibinfo{pages}{012003} (\bibinfo{year}{2003}), \eprint{hep-ph/0305035}.

\bibitem[{\citenamefont{Dai et~al.}(2003)\citenamefont{Dai, Huang, Liu, and
  Zhu}}]{Dai:2003yg}
\bibinfo{author}{\bibfnamefont{Y.-B.} \bibnamefont{Dai}},
  \bibinfo{author}{\bibfnamefont{C.-S.} \bibnamefont{Huang}},
  \bibinfo{author}{\bibfnamefont{C.}~\bibnamefont{Liu}}, \bibnamefont{and}
  \bibinfo{author}{\bibfnamefont{S.-L.} \bibnamefont{Zhu}},
  \bibinfo{journal}{Phys. Rev.} \textbf{\bibinfo{volume}{D68}},
  \bibinfo{pages}{114011} (\bibinfo{year}{2003}), \eprint{hep-ph/0306274}.

\bibitem[{\citenamefont{Narison}(2005)}]{Narison:2003td}
\bibinfo{author}{\bibfnamefont{S.}~\bibnamefont{Narison}},
  \bibinfo{journal}{Phys. Lett.} \textbf{\bibinfo{volume}{B605}},
  \bibinfo{pages}{319} (\bibinfo{year}{2005}), \eprint{hep-ph/0307248}.

\bibitem[{\citenamefont{Szczepaniak}(2003)}]{Szczepaniak:2003vy}
\bibinfo{author}{\bibfnamefont{A.~P.} \bibnamefont{Szczepaniak}},
  \bibinfo{journal}{Phys. Lett.} \textbf{\bibinfo{volume}{B567}},
  \bibinfo{pages}{23} (\bibinfo{year}{2003}), \eprint{hep-ph/0305060}.

\bibitem[{\citenamefont{Browder et~al.}(2004)\citenamefont{Browder, Pakvasa,
  and Petrov}}]{Browder:2003fk}
\bibinfo{author}{\bibfnamefont{T.~E.} \bibnamefont{Browder}},
  \bibinfo{author}{\bibfnamefont{S.}~\bibnamefont{Pakvasa}}, \bibnamefont{and}
  \bibinfo{author}{\bibfnamefont{A.~A.} \bibnamefont{Petrov}},
  \bibinfo{journal}{Phys. Lett.} \textbf{\bibinfo{volume}{B578}},
  \bibinfo{pages}{365} (\bibinfo{year}{2004}), \eprint{hep-ph/0307054}.

\bibitem[{\citenamefont{Barnes et~al.}(2003)\citenamefont{Barnes, Close, and
  Lipkin}}]{Barnes:2003dj}
\bibinfo{author}{\bibfnamefont{T.}~\bibnamefont{Barnes}},
  \bibinfo{author}{\bibfnamefont{F.~E.} \bibnamefont{Close}}, \bibnamefont{and}
  \bibinfo{author}{\bibfnamefont{H.~J.} \bibnamefont{Lipkin}},
  \bibinfo{journal}{Phys. Rev.} \textbf{\bibinfo{volume}{D68}},
  \bibinfo{pages}{054006} (\bibinfo{year}{2003}), \eprint{hep-ph/0305025}.

\bibitem[{\citenamefont{Cheng and Hou}(2003)}]{Cheng:2003kg}
\bibinfo{author}{\bibfnamefont{H.-Y.} \bibnamefont{Cheng}} \bibnamefont{and}
  \bibinfo{author}{\bibfnamefont{W.-S.} \bibnamefont{Hou}},
  \bibinfo{journal}{Phys. Lett.} \textbf{\bibinfo{volume}{B566}},
  \bibinfo{pages}{193} (\bibinfo{year}{2003}), \eprint{hep-ph/0305038}.

\bibitem[{\citenamefont{Chen and Li}(2004)}]{Chen:2004dy}
\bibinfo{author}{\bibfnamefont{Y.-Q.} \bibnamefont{Chen}} \bibnamefont{and}
  \bibinfo{author}{\bibfnamefont{X.-Q.} \bibnamefont{Li}},
  \bibinfo{journal}{Phys. Rev. Lett.} \textbf{\bibinfo{volume}{93}},
  \bibinfo{pages}{232001} (\bibinfo{year}{2004}), \eprint{hep-ph/0407062}.

\bibitem[{\citenamefont{Dmitrasinovic}(2005)}]{Dmitrasinovic:2005gc}
\bibinfo{author}{\bibfnamefont{V.}~\bibnamefont{Dmitrasinovic}},
  \bibinfo{journal}{Phys. Rev. Lett.} \textbf{\bibinfo{volume}{94}},
  \bibinfo{pages}{162002} (\bibinfo{year}{2005}).

\bibitem[{\citenamefont{Zhang}(2019)}]{Zhang:2018mnm}
\bibinfo{author}{\bibfnamefont{J.-R.} \bibnamefont{Zhang}},
  \bibinfo{journal}{Phys. Lett.} \textbf{\bibinfo{volume}{B789}},
  \bibinfo{pages}{432} (\bibinfo{year}{2019}), \eprint{1801.08725}.

\bibitem[{\citenamefont{Terasaki}(2003)}]{Terasaki:2003qa}
\bibinfo{author}{\bibfnamefont{K.}~\bibnamefont{Terasaki}},
  \bibinfo{journal}{Phys. Rev.} \textbf{\bibinfo{volume}{D68}},
  \bibinfo{pages}{011501} (\bibinfo{year}{2003}), \eprint{hep-ph/0305213}.

\bibitem[{\citenamefont{Maiani et~al.}(2005)\citenamefont{Maiani, Piccinini,
  Polosa, and Riquer}}]{Maiani:2004vq}
\bibinfo{author}{\bibfnamefont{L.}~\bibnamefont{Maiani}},
  \bibinfo{author}{\bibfnamefont{F.}~\bibnamefont{Piccinini}},
  \bibinfo{author}{\bibfnamefont{A.}~\bibnamefont{Polosa}}, \bibnamefont{and}
  \bibinfo{author}{\bibfnamefont{V.}~\bibnamefont{Riquer}},
  \bibinfo{journal}{Phys.Rev.} \textbf{\bibinfo{volume}{D71}},
  \bibinfo{pages}{014028} (\bibinfo{year}{2005}), \eprint{hep-ph/0412098}.

\bibitem[{\citenamefont{Kolomeitsev and Lutz}(2004)}]{Kolomeitsev:2003ac}
\bibinfo{author}{\bibfnamefont{E.~E.} \bibnamefont{Kolomeitsev}}
  \bibnamefont{and} \bibinfo{author}{\bibfnamefont{M.~F.~M.}
  \bibnamefont{Lutz}}, \bibinfo{journal}{Phys. Lett.}
  \textbf{\bibinfo{volume}{B582}}, \bibinfo{pages}{39} (\bibinfo{year}{2004}),
  \eprint{hep-ph/0307133}.

\bibitem[{\citenamefont{Hofmann and Lutz}(2004)}]{Hofmann:2003je}
\bibinfo{author}{\bibfnamefont{J.}~\bibnamefont{Hofmann}} \bibnamefont{and}
  \bibinfo{author}{\bibfnamefont{M.~F.~M.} \bibnamefont{Lutz}},
  \bibinfo{journal}{Nucl. Phys.} \textbf{\bibinfo{volume}{A733}},
  \bibinfo{pages}{142} (\bibinfo{year}{2004}), \eprint{hep-ph/0308263}.

\bibitem[{\citenamefont{Guo et~al.}(2006)\citenamefont{Guo, Shen, Chiang, Ping,
  and Zou}}]{Guo:2006fu}
\bibinfo{author}{\bibfnamefont{F.-K.} \bibnamefont{Guo}},
  \bibinfo{author}{\bibfnamefont{P.-N.} \bibnamefont{Shen}},
  \bibinfo{author}{\bibfnamefont{H.-C.} \bibnamefont{Chiang}},
  \bibinfo{author}{\bibfnamefont{R.-G.} \bibnamefont{Ping}}, \bibnamefont{and}
  \bibinfo{author}{\bibfnamefont{B.-S.} \bibnamefont{Zou}},
  \bibinfo{journal}{Phys. Lett.} \textbf{\bibinfo{volume}{B641}},
  \bibinfo{pages}{278} (\bibinfo{year}{2006}), \eprint{hep-ph/0603072}.

\bibitem[{\citenamefont{Gamermann et~al.}(2007)\citenamefont{Gamermann, Oset,
  Strottman, and Vicente~Vacas}}]{Gamermann:2006nm}
\bibinfo{author}{\bibfnamefont{D.}~\bibnamefont{Gamermann}},
  \bibinfo{author}{\bibfnamefont{E.}~\bibnamefont{Oset}},
  \bibinfo{author}{\bibfnamefont{D.}~\bibnamefont{Strottman}},
  \bibnamefont{and} \bibinfo{author}{\bibfnamefont{M.~J.}
  \bibnamefont{Vicente~Vacas}}, \bibinfo{journal}{Phys. Rev.}
  \textbf{\bibinfo{volume}{D76}}, \bibinfo{pages}{074016}
  (\bibinfo{year}{2007}), \eprint{hep-ph/0612179}.

\bibitem[{\citenamefont{Guo et~al.}(2008)\citenamefont{Guo, Hanhart, Krewald,
  and Meissner}}]{Guo:2008gp}
\bibinfo{author}{\bibfnamefont{F.-K.} \bibnamefont{Guo}},
  \bibinfo{author}{\bibfnamefont{C.}~\bibnamefont{Hanhart}},
  \bibinfo{author}{\bibfnamefont{S.}~\bibnamefont{Krewald}}, \bibnamefont{and}
  \bibinfo{author}{\bibfnamefont{U.-G.} \bibnamefont{Meissner}},
  \bibinfo{journal}{Phys. Lett.} \textbf{\bibinfo{volume}{B666}},
  \bibinfo{pages}{251} (\bibinfo{year}{2008}), \eprint{0806.3374}.

\bibitem[{\citenamefont{Guo et~al.}(2009)\citenamefont{Guo, Hanhart, and
  Meissner}}]{Guo:2009ct}
\bibinfo{author}{\bibfnamefont{F.-K.} \bibnamefont{Guo}},
  \bibinfo{author}{\bibfnamefont{C.}~\bibnamefont{Hanhart}}, \bibnamefont{and}
  \bibinfo{author}{\bibfnamefont{U.-G.} \bibnamefont{Meissner}},
  \bibinfo{journal}{Eur. Phys. J.} \textbf{\bibinfo{volume}{A40}},
  \bibinfo{pages}{171} (\bibinfo{year}{2009}), \eprint{0901.1597}.

\bibitem[{\citenamefont{Cleven et~al.}(2011)\citenamefont{Cleven, Guo, Hanhart,
  and Meissner}}]{Cleven:2010aw}
\bibinfo{author}{\bibfnamefont{M.}~\bibnamefont{Cleven}},
  \bibinfo{author}{\bibfnamefont{F.-K.} \bibnamefont{Guo}},
  \bibinfo{author}{\bibfnamefont{C.}~\bibnamefont{Hanhart}}, \bibnamefont{and}
  \bibinfo{author}{\bibfnamefont{U.-G.} \bibnamefont{Meissner}},
  \bibinfo{journal}{Eur. Phys. J.} \textbf{\bibinfo{volume}{A47}},
  \bibinfo{pages}{19} (\bibinfo{year}{2011}), \eprint{1009.3804}.

\bibitem[{\citenamefont{Martinez~Torres
  et~al.}(2012)\citenamefont{Martinez~Torres, Dai, Koren, Jido, and
  Oset}}]{MartinezTorres:2011pr}
\bibinfo{author}{\bibfnamefont{A.}~\bibnamefont{Martinez~Torres}},
  \bibinfo{author}{\bibfnamefont{L.~R.} \bibnamefont{Dai}},
  \bibinfo{author}{\bibfnamefont{C.}~\bibnamefont{Koren}},
  \bibinfo{author}{\bibfnamefont{D.}~\bibnamefont{Jido}}, \bibnamefont{and}
  \bibinfo{author}{\bibfnamefont{E.}~\bibnamefont{Oset}},
  \bibinfo{journal}{Phys. Rev.} \textbf{\bibinfo{volume}{D85}},
  \bibinfo{pages}{014027} (\bibinfo{year}{2012}), \eprint{1109.0396}.

\bibitem[{\citenamefont{Yao et~al.}(2015)\citenamefont{Yao, Du, Guo, and
  Mei{\ss}ner}}]{Yao:2015qia}
\bibinfo{author}{\bibfnamefont{D.-L.} \bibnamefont{Yao}},
  \bibinfo{author}{\bibfnamefont{M.-L.} \bibnamefont{Du}},
  \bibinfo{author}{\bibfnamefont{F.-K.} \bibnamefont{Guo}}, \bibnamefont{and}
  \bibinfo{author}{\bibfnamefont{U.-G.} \bibnamefont{Mei{\ss}ner}},
  \bibinfo{journal}{JHEP} \textbf{\bibinfo{volume}{11}}, \bibinfo{pages}{058}
  (\bibinfo{year}{2015}), \eprint{1502.05981}.

\bibitem[{\citenamefont{Guo et~al.}(2015)\citenamefont{Guo, Mei{\ss}ner, and
  Yao}}]{Guo:2015dha}
\bibinfo{author}{\bibfnamefont{Z.-H.} \bibnamefont{Guo}},
  \bibinfo{author}{\bibfnamefont{U.-G.} \bibnamefont{Mei{\ss}ner}},
  \bibnamefont{and} \bibinfo{author}{\bibfnamefont{D.-L.} \bibnamefont{Yao}},
  \bibinfo{journal}{Phys. Rev.} \textbf{\bibinfo{volume}{D92}},
  \bibinfo{pages}{094008} (\bibinfo{year}{2015}), \eprint{1507.03123}.

\bibitem[{\citenamefont{Albaladejo et~al.}(2017)\citenamefont{Albaladejo,
  Fernandez-Soler, Guo, and Nieves}}]{Albaladejo:2016lbb}
\bibinfo{author}{\bibfnamefont{M.}~\bibnamefont{Albaladejo}},
  \bibinfo{author}{\bibfnamefont{P.}~\bibnamefont{Fernandez-Soler}},
  \bibinfo{author}{\bibfnamefont{F.-K.} \bibnamefont{Guo}}, \bibnamefont{and}
  \bibinfo{author}{\bibfnamefont{J.}~\bibnamefont{Nieves}},
  \bibinfo{journal}{Phys. Lett.} \textbf{\bibinfo{volume}{B767}},
  \bibinfo{pages}{465} (\bibinfo{year}{2017}), \eprint{1610.06727}.

\bibitem[{\citenamefont{Du et~al.}(2017)\citenamefont{Du, Guo, Mei{\ss}ner, and
  Yao}}]{Du:2017ttu}
\bibinfo{author}{\bibfnamefont{M.-L.} \bibnamefont{Du}},
  \bibinfo{author}{\bibfnamefont{F.-K.} \bibnamefont{Guo}},
  \bibinfo{author}{\bibfnamefont{U.-G.} \bibnamefont{Mei{\ss}ner}},
  \bibnamefont{and} \bibinfo{author}{\bibfnamefont{D.-L.} \bibnamefont{Yao}},
  \bibinfo{journal}{Eur. Phys. J.} \textbf{\bibinfo{volume}{C77}},
  \bibinfo{pages}{728} (\bibinfo{year}{2017}), \eprint{1703.10836}.

\bibitem[{\citenamefont{Guo et~al.}(2018)\citenamefont{Guo, Heo, and
  Lutz}}]{Guo:2018kno}
\bibinfo{author}{\bibfnamefont{X.-Y.} \bibnamefont{Guo}},
  \bibinfo{author}{\bibfnamefont{Y.}~\bibnamefont{Heo}}, \bibnamefont{and}
  \bibinfo{author}{\bibfnamefont{M.~F.~M.} \bibnamefont{Lutz}},
  \bibinfo{journal}{Phys. Rev.} \textbf{\bibinfo{volume}{D98}},
  \bibinfo{pages}{014510} (\bibinfo{year}{2018}), \eprint{1801.10122}.

\bibitem[{\citenamefont{Albaladejo et~al.}(2018)\citenamefont{Albaladejo,
  Fernandez-Soler, Nieves, and Ortega}}]{Albaladejo:2018mhb}
\bibinfo{author}{\bibfnamefont{M.}~\bibnamefont{Albaladejo}},
  \bibinfo{author}{\bibfnamefont{P.}~\bibnamefont{Fernandez-Soler}},
  \bibinfo{author}{\bibfnamefont{J.}~\bibnamefont{Nieves}}, \bibnamefont{and}
  \bibinfo{author}{\bibfnamefont{P.~G.} \bibnamefont{Ortega}},
  \bibinfo{journal}{Eur. Phys. J.} \textbf{\bibinfo{volume}{C78}},
  \bibinfo{pages}{722} (\bibinfo{year}{2018}), \eprint{1805.07104}.

\bibitem[{\citenamefont{Altenbuchinger and
  Geng}(2014)}]{Altenbuchinger:2013gaa}
\bibinfo{author}{\bibfnamefont{M.}~\bibnamefont{Altenbuchinger}}
  \bibnamefont{and} \bibinfo{author}{\bibfnamefont{L.-S.} \bibnamefont{Geng}},
  \bibinfo{journal}{Phys. Rev.} \textbf{\bibinfo{volume}{D89}},
  \bibinfo{pages}{054008} (\bibinfo{year}{2014}), \eprint{1310.5224}.

\bibitem[{\citenamefont{Altenbuchinger
  et~al.}(2014)\citenamefont{Altenbuchinger, Geng, and
  Weise}}]{Altenbuchinger:2013vwa}
\bibinfo{author}{\bibfnamefont{M.}~\bibnamefont{Altenbuchinger}},
  \bibinfo{author}{\bibfnamefont{L.~S.} \bibnamefont{Geng}}, \bibnamefont{and}
  \bibinfo{author}{\bibfnamefont{W.}~\bibnamefont{Weise}},
  \bibinfo{journal}{Phys. Rev.} \textbf{\bibinfo{volume}{D89}},
  \bibinfo{pages}{014026} (\bibinfo{year}{2014}), \eprint{1309.4743}.

\bibitem[{\citenamefont{Geng et~al.}(2010)\citenamefont{Geng, Kaiser,
  Martin-Camalich, and Weise}}]{Geng:2010vw}
\bibinfo{author}{\bibfnamefont{L.~S.} \bibnamefont{Geng}},
  \bibinfo{author}{\bibfnamefont{N.}~\bibnamefont{Kaiser}},
  \bibinfo{author}{\bibfnamefont{J.}~\bibnamefont{Martin-Camalich}},
  \bibnamefont{and} \bibinfo{author}{\bibfnamefont{W.}~\bibnamefont{Weise}},
  \bibinfo{journal}{Phys. Rev.} \textbf{\bibinfo{volume}{D82}},
  \bibinfo{pages}{054022} (\bibinfo{year}{2010}), \eprint{1008.0383}.

\bibitem[{\citenamefont{Wang and Wang}(2012)}]{Wang:2012bu}
\bibinfo{author}{\bibfnamefont{P.}~\bibnamefont{Wang}} \bibnamefont{and}
  \bibinfo{author}{\bibfnamefont{X.~G.} \bibnamefont{Wang}},
  \bibinfo{journal}{Phys. Rev.} \textbf{\bibinfo{volume}{D86}},
  \bibinfo{pages}{014030} (\bibinfo{year}{2012}), \eprint{1204.5553}.

\bibitem[{\citenamefont{Liu et~al.}(2009)\citenamefont{Liu, Liu, and
  Zhu}}]{Liu:2009uz}
\bibinfo{author}{\bibfnamefont{Y.-R.} \bibnamefont{Liu}},
  \bibinfo{author}{\bibfnamefont{X.}~\bibnamefont{Liu}}, \bibnamefont{and}
  \bibinfo{author}{\bibfnamefont{S.-L.} \bibnamefont{Zhu}},
  \bibinfo{journal}{Phys. Rev.} \textbf{\bibinfo{volume}{D79}},
  \bibinfo{pages}{094026} (\bibinfo{year}{2009}), \eprint{0904.1770}.

\bibitem[{\citenamefont{Guo et~al.}(2019)\citenamefont{Guo, Liu, Mei{\ss}ner,
  Oller, and Rusetsky}}]{Guo:2018tjx}
\bibinfo{author}{\bibfnamefont{Z.-H.} \bibnamefont{Guo}},
  \bibinfo{author}{\bibfnamefont{L.}~\bibnamefont{Liu}},
  \bibinfo{author}{\bibfnamefont{U.-G.} \bibnamefont{Mei{\ss}ner}},
  \bibinfo{author}{\bibfnamefont{J.~A.} \bibnamefont{Oller}}, \bibnamefont{and}
  \bibinfo{author}{\bibfnamefont{A.}~\bibnamefont{Rusetsky}},
  \bibinfo{journal}{Eur. Phys. J.} \textbf{\bibinfo{volume}{C79}},
  \bibinfo{pages}{13} (\bibinfo{year}{2019}), \eprint{1811.05585}.

\bibitem[{\citenamefont{Liu et~al.}(2013)\citenamefont{Liu, Orginos, Guo,
  Hanhart, and Meissner}}]{Liu:2012zya}
\bibinfo{author}{\bibfnamefont{L.}~\bibnamefont{Liu}},
  \bibinfo{author}{\bibfnamefont{K.}~\bibnamefont{Orginos}},
  \bibinfo{author}{\bibfnamefont{F.-K.} \bibnamefont{Guo}},
  \bibinfo{author}{\bibfnamefont{C.}~\bibnamefont{Hanhart}}, \bibnamefont{and}
  \bibinfo{author}{\bibfnamefont{U.-G.} \bibnamefont{Meissner}},
  \bibinfo{journal}{Phys. Rev.} \textbf{\bibinfo{volume}{D87}},
  \bibinfo{pages}{014508} (\bibinfo{year}{2013}), \eprint{1208.4535}.

\bibitem[{\citenamefont{Mohler et~al.}(2013)\citenamefont{Mohler, Lang,
  Leskovec, Prelovsek, and Woloshyn}}]{Mohler:2013rwa}
\bibinfo{author}{\bibfnamefont{D.}~\bibnamefont{Mohler}},
  \bibinfo{author}{\bibfnamefont{C.~B.} \bibnamefont{Lang}},
  \bibinfo{author}{\bibfnamefont{L.}~\bibnamefont{Leskovec}},
  \bibinfo{author}{\bibfnamefont{S.}~\bibnamefont{Prelovsek}},
  \bibnamefont{and} \bibinfo{author}{\bibfnamefont{R.~M.}
  \bibnamefont{Woloshyn}}, \bibinfo{journal}{Phys. Rev. Lett.}
  \textbf{\bibinfo{volume}{111}}, \bibinfo{pages}{222001}
  (\bibinfo{year}{2013}), \eprint{1308.3175}.

\bibitem[{\citenamefont{Lang et~al.}(2014)\citenamefont{Lang, Leskovec, Mohler,
  Prelovsek, and Woloshyn}}]{Lang:2014yfa}
\bibinfo{author}{\bibfnamefont{C.~B.} \bibnamefont{Lang}},
  \bibinfo{author}{\bibfnamefont{L.}~\bibnamefont{Leskovec}},
  \bibinfo{author}{\bibfnamefont{D.}~\bibnamefont{Mohler}},
  \bibinfo{author}{\bibfnamefont{S.}~\bibnamefont{Prelovsek}},
  \bibnamefont{and} \bibinfo{author}{\bibfnamefont{R.~M.}
  \bibnamefont{Woloshyn}}, \bibinfo{journal}{Phys. Rev.}
  \textbf{\bibinfo{volume}{D90}}, \bibinfo{pages}{034510}
  (\bibinfo{year}{2014}), \eprint{1403.8103}.

\bibitem[{\citenamefont{Mart{\'\i}nez~Torres
  et~al.}(2015)\citenamefont{Mart{\'\i}nez~Torres, Oset, Prelovsek, and
  Ramos}}]{Torres:2014vna}
\bibinfo{author}{\bibfnamefont{A.}~\bibnamefont{Mart{\'\i}nez~Torres}},
  \bibinfo{author}{\bibfnamefont{E.}~\bibnamefont{Oset}},
  \bibinfo{author}{\bibfnamefont{S.}~\bibnamefont{Prelovsek}},
  \bibnamefont{and} \bibinfo{author}{\bibfnamefont{A.}~\bibnamefont{Ramos}},
  \bibinfo{journal}{JHEP} \textbf{\bibinfo{volume}{05}}, \bibinfo{pages}{153}
  (\bibinfo{year}{2015}), \eprint{1412.1706}.

\bibitem[{\citenamefont{Bali et~al.}(2017)\citenamefont{Bali, Collins, Cox, and
  Sch{\"a}fer}}]{Bali:2017pdv}
\bibinfo{author}{\bibfnamefont{G.~S.} \bibnamefont{Bali}},
  \bibinfo{author}{\bibfnamefont{S.}~\bibnamefont{Collins}},
  \bibinfo{author}{\bibfnamefont{A.}~\bibnamefont{Cox}}, \bibnamefont{and}
  \bibinfo{author}{\bibfnamefont{A.}~\bibnamefont{Sch{\"a}fer}},
  \bibinfo{journal}{Phys. Rev.} \textbf{\bibinfo{volume}{D96}},
  \bibinfo{pages}{074501} (\bibinfo{year}{2017}), \eprint{1706.01247}.

\bibitem[{\citenamefont{Guo}(2019)}]{Guo:2019dpg}
\bibinfo{author}{\bibfnamefont{F.-K.} \bibnamefont{Guo}}, \bibinfo{journal}{EPJ
  Web Conf.} \textbf{\bibinfo{volume}{202}}, \bibinfo{pages}{02001}
  (\bibinfo{year}{2019}).

\bibitem[{\citenamefont{Sanchez~Sanchez
  et~al.}(2018)\citenamefont{Sanchez~Sanchez, Geng, Lu, Hyodo, and
  Valderrama}}]{SanchezSanchez:2017xtl}
\bibinfo{author}{\bibfnamefont{M.}~\bibnamefont{Sanchez~Sanchez}},
  \bibinfo{author}{\bibfnamefont{L.-S.} \bibnamefont{Geng}},
  \bibinfo{author}{\bibfnamefont{J.-X.} \bibnamefont{Lu}},
  \bibinfo{author}{\bibfnamefont{T.}~\bibnamefont{Hyodo}}, \bibnamefont{and}
  \bibinfo{author}{\bibfnamefont{M.~P.} \bibnamefont{Valderrama}},
  \bibinfo{journal}{Phys. Rev.} \textbf{\bibinfo{volume}{D98}},
  \bibinfo{pages}{054001} (\bibinfo{year}{2018}), \eprint{1707.03802}.

\bibitem[{\citenamefont{Martinez~Torres
  et~al.}(2019)\citenamefont{Martinez~Torres, Khemchandani, and
  Geng}}]{MartinezTorres:2018zbl}
\bibinfo{author}{\bibfnamefont{A.}~\bibnamefont{Martinez~Torres}},
  \bibinfo{author}{\bibfnamefont{K.}~\bibnamefont{Khemchandani}},
  \bibnamefont{and} \bibinfo{author}{\bibfnamefont{L.-S.} \bibnamefont{Geng}},
  \bibinfo{journal}{Phys. Rev.} \textbf{\bibinfo{volume}{D99}},
  \bibinfo{pages}{076017} (\bibinfo{year}{2019}), \eprint{1809.01059}.

\bibitem[{\citenamefont{Wu et~al.}(2019)\citenamefont{Wu, Liu, Geng, Hiyama,
  and Valderrama}}]{Wu:2019vsy}
\bibinfo{author}{\bibfnamefont{T.-W.} \bibnamefont{Wu}},
  \bibinfo{author}{\bibfnamefont{M.-Z.} \bibnamefont{Liu}},
  \bibinfo{author}{\bibfnamefont{L.-S.} \bibnamefont{Geng}},
  \bibinfo{author}{\bibfnamefont{E.}~\bibnamefont{Hiyama}}, \bibnamefont{and}
  \bibinfo{author}{\bibfnamefont{M.~P.} \bibnamefont{Valderrama}}
  (\bibinfo{year}{2019}), \eprint{1906.11995}.

\bibitem[{\citenamefont{Ma et~al.}(2019)\citenamefont{Ma, Wang, and
  Mei{\ss}ner}}]{Ma:2017ery}
\bibinfo{author}{\bibfnamefont{L.}~\bibnamefont{Ma}},
  \bibinfo{author}{\bibfnamefont{Q.}~\bibnamefont{Wang}}, \bibnamefont{and}
  \bibinfo{author}{\bibfnamefont{U.-G.} \bibnamefont{Mei{\ss}ner}},
  \bibinfo{journal}{Chin. Phys.} \textbf{\bibinfo{volume}{C43}},
  \bibinfo{pages}{014102} (\bibinfo{year}{2019}), \eprint{1711.06143}.

\bibitem[{\citenamefont{Ren et~al.}(2018)\citenamefont{Ren, Malabarba, Geng,
  Khemchandani, and Mart{\'\i}nez~Torres}}]{Ren:2018pcd}
\bibinfo{author}{\bibfnamefont{X.-L.} \bibnamefont{Ren}},
  \bibinfo{author}{\bibfnamefont{B.~B.} \bibnamefont{Malabarba}},
  \bibinfo{author}{\bibfnamefont{L.-S.} \bibnamefont{Geng}},
  \bibinfo{author}{\bibfnamefont{K.~P.} \bibnamefont{Khemchandani}},
  \bibnamefont{and}
  \bibinfo{author}{\bibfnamefont{A.}~\bibnamefont{Mart{\'\i}nez~Torres}},
  \bibinfo{journal}{Phys. Lett.} \textbf{\bibinfo{volume}{B785}},
  \bibinfo{pages}{112} (\bibinfo{year}{2018}), \eprint{1805.08330}.

\bibitem[{\citenamefont{Ren et~al.}(2019)\citenamefont{Ren, Malabarba,
  Khemchandani, and Martinez~Torres}}]{Ren:2019umd}
\bibinfo{author}{\bibfnamefont{X.-L.} \bibnamefont{Ren}},
  \bibinfo{author}{\bibfnamefont{B.~B.} \bibnamefont{Malabarba}},
  \bibinfo{author}{\bibfnamefont{K.~P.} \bibnamefont{Khemchandani}},
  \bibnamefont{and}
  \bibinfo{author}{\bibfnamefont{A.}~\bibnamefont{Martinez~Torres}},
  \bibinfo{journal}{JHEP} \textbf{\bibinfo{volume}{05}}, \bibinfo{pages}{103}
  (\bibinfo{year}{2019}), \eprint{1904.06768}.

\bibitem[{\citenamefont{Debastiani et~al.}(2017)\citenamefont{Debastiani, Dias,
  and Oset}}]{Debastiani:2017vhv}
\bibinfo{author}{\bibfnamefont{V.~R.} \bibnamefont{Debastiani}},
  \bibinfo{author}{\bibfnamefont{J.~M.} \bibnamefont{Dias}}, \bibnamefont{and}
  \bibinfo{author}{\bibfnamefont{E.}~\bibnamefont{Oset}},
  \bibinfo{journal}{Phys. Rev.} \textbf{\bibinfo{volume}{D96}},
  \bibinfo{pages}{016014} (\bibinfo{year}{2017}), \eprint{1705.09257}.

\bibitem[{\citenamefont{Yamagata-Sekihara and
  Sekihara}(2019)}]{Yamagata-Sekihara:2018gah}
\bibinfo{author}{\bibfnamefont{J.}~\bibnamefont{Yamagata-Sekihara}}
  \bibnamefont{and} \bibinfo{author}{\bibfnamefont{T.}~\bibnamefont{Sekihara}},
  \bibinfo{journal}{Phys. Rev.} \textbf{\bibinfo{volume}{C100}},
  \bibinfo{pages}{015203} (\bibinfo{year}{2019}), \eprint{1809.08896}.

\bibitem[{\citenamefont{Faessler
  et~al.}(2007{\natexlab{a}})\citenamefont{Faessler, Gutsche, Lyubovitskij, and
  Ma}}]{Faessler:2007gv}
\bibinfo{author}{\bibfnamefont{A.}~\bibnamefont{Faessler}},
  \bibinfo{author}{\bibfnamefont{T.}~\bibnamefont{Gutsche}},
  \bibinfo{author}{\bibfnamefont{V.~E.} \bibnamefont{Lyubovitskij}},
  \bibnamefont{and} \bibinfo{author}{\bibfnamefont{Y.-L.} \bibnamefont{Ma}},
  \bibinfo{journal}{Phys. Rev.} \textbf{\bibinfo{volume}{D76}},
  \bibinfo{pages}{014005} (\bibinfo{year}{2007}{\natexlab{a}}),
  \eprint{0705.0254}.

\bibitem[{\citenamefont{Faessler
  et~al.}(2007{\natexlab{b}})\citenamefont{Faessler, Gutsche, Lyubovitskij, and
  Ma}}]{Faessler:2007us}
\bibinfo{author}{\bibfnamefont{A.}~\bibnamefont{Faessler}},
  \bibinfo{author}{\bibfnamefont{T.}~\bibnamefont{Gutsche}},
  \bibinfo{author}{\bibfnamefont{V.~E.} \bibnamefont{Lyubovitskij}},
  \bibnamefont{and} \bibinfo{author}{\bibfnamefont{Y.-L.} \bibnamefont{Ma}},
  \bibinfo{journal}{Phys. Rev.} \textbf{\bibinfo{volume}{D76}},
  \bibinfo{pages}{114008} (\bibinfo{year}{2007}{\natexlab{b}}),
  \eprint{0709.3946}.

\bibitem[{\citenamefont{Dong et~al.}(2008)\citenamefont{Dong, Faessler,
  Gutsche, and Lyubovitskij}}]{Dong:2008gb}
\bibinfo{author}{\bibfnamefont{Y.-b.} \bibnamefont{Dong}},
  \bibinfo{author}{\bibfnamefont{A.}~\bibnamefont{Faessler}},
  \bibinfo{author}{\bibfnamefont{T.}~\bibnamefont{Gutsche}}, \bibnamefont{and}
  \bibinfo{author}{\bibfnamefont{V.~E.} \bibnamefont{Lyubovitskij}},
  \bibinfo{journal}{Phys. Rev.} \textbf{\bibinfo{volume}{D77}},
  \bibinfo{pages}{094013} (\bibinfo{year}{2008}), \eprint{0802.3610}.

\bibitem[{\citenamefont{Dong et~al.}(2011{\natexlab{a}})\citenamefont{Dong,
  Faessler, Gutsche, and Lyubovitskij}}]{Dong:2009uf}
\bibinfo{author}{\bibfnamefont{Y.}~\bibnamefont{Dong}},
  \bibinfo{author}{\bibfnamefont{A.}~\bibnamefont{Faessler}},
  \bibinfo{author}{\bibfnamefont{T.}~\bibnamefont{Gutsche}}, \bibnamefont{and}
  \bibinfo{author}{\bibfnamefont{V.~E.} \bibnamefont{Lyubovitskij}},
  \bibinfo{journal}{J. Phys.} \textbf{\bibinfo{volume}{G38}},
  \bibinfo{pages}{015001} (\bibinfo{year}{2011}{\natexlab{a}}),
  \eprint{0909.0380}.

\bibitem[{\citenamefont{Dong et~al.}(2009)\citenamefont{Dong, Faessler,
  Gutsche, Kovalenko, and Lyubovitskij}}]{Dong:2009yp}
\bibinfo{author}{\bibfnamefont{Y.}~\bibnamefont{Dong}},
  \bibinfo{author}{\bibfnamefont{A.}~\bibnamefont{Faessler}},
  \bibinfo{author}{\bibfnamefont{T.}~\bibnamefont{Gutsche}},
  \bibinfo{author}{\bibfnamefont{S.}~\bibnamefont{Kovalenko}},
  \bibnamefont{and} \bibinfo{author}{\bibfnamefont{V.~E.}
  \bibnamefont{Lyubovitskij}}, \bibinfo{journal}{Phys. Rev.}
  \textbf{\bibinfo{volume}{D79}}, \bibinfo{pages}{094013}
  (\bibinfo{year}{2009}), \eprint{0903.5416}.

\bibitem[{\citenamefont{Dong et~al.}(2017{\natexlab{a}})\citenamefont{Dong,
  Faessler, Gutsche, L{\"u}, and Lyubovitskij}}]{Dong:2017rmg}
\bibinfo{author}{\bibfnamefont{Y.}~\bibnamefont{Dong}},
  \bibinfo{author}{\bibfnamefont{A.}~\bibnamefont{Faessler}},
  \bibinfo{author}{\bibfnamefont{T.}~\bibnamefont{Gutsche}},
  \bibinfo{author}{\bibfnamefont{Q.}~\bibnamefont{L{\"u}}}, \bibnamefont{and}
  \bibinfo{author}{\bibfnamefont{V.~E.} \bibnamefont{Lyubovitskij}},
  \bibinfo{journal}{Phys. Rev.} \textbf{\bibinfo{volume}{D96}},
  \bibinfo{pages}{074027} (\bibinfo{year}{2017}{\natexlab{a}}),
  \eprint{1705.09631}.

\bibitem[{\citenamefont{Dong et~al.}(2014{\natexlab{a}})\citenamefont{Dong,
  Faessler, Gutsche, and Lyubovitskij}}]{Dong:2014ksa}
\bibinfo{author}{\bibfnamefont{Y.}~\bibnamefont{Dong}},
  \bibinfo{author}{\bibfnamefont{A.}~\bibnamefont{Faessler}},
  \bibinfo{author}{\bibfnamefont{T.}~\bibnamefont{Gutsche}}, \bibnamefont{and}
  \bibinfo{author}{\bibfnamefont{V.~E.} \bibnamefont{Lyubovitskij}},
  \bibinfo{journal}{Phys. Rev.} \textbf{\bibinfo{volume}{D90}},
  \bibinfo{pages}{094001} (\bibinfo{year}{2014}{\natexlab{a}}),
  \eprint{1407.3949}.

\bibitem[{\citenamefont{Dong et~al.}(2014{\natexlab{b}})\citenamefont{Dong,
  Faessler, Gutsche, and Lyubovitskij}}]{Dong:2014zka}
\bibinfo{author}{\bibfnamefont{Y.}~\bibnamefont{Dong}},
  \bibinfo{author}{\bibfnamefont{A.}~\bibnamefont{Faessler}},
  \bibinfo{author}{\bibfnamefont{T.}~\bibnamefont{Gutsche}}, \bibnamefont{and}
  \bibinfo{author}{\bibfnamefont{V.~E.} \bibnamefont{Lyubovitskij}},
  \bibinfo{journal}{Phys. Rev.} \textbf{\bibinfo{volume}{D90}},
  \bibinfo{pages}{074032} (\bibinfo{year}{2014}{\natexlab{b}}),
  \eprint{1404.6161}.

\bibitem[{\citenamefont{Dong et~al.}(2014{\natexlab{c}})\citenamefont{Dong,
  Faessler, Gutsche, and Lyubovitskij}}]{Dong:2013kta}
\bibinfo{author}{\bibfnamefont{Y.}~\bibnamefont{Dong}},
  \bibinfo{author}{\bibfnamefont{A.}~\bibnamefont{Faessler}},
  \bibinfo{author}{\bibfnamefont{T.}~\bibnamefont{Gutsche}}, \bibnamefont{and}
  \bibinfo{author}{\bibfnamefont{V.~E.} \bibnamefont{Lyubovitskij}},
  \bibinfo{journal}{Phys. Rev.} \textbf{\bibinfo{volume}{D89}},
  \bibinfo{pages}{034018} (\bibinfo{year}{2014}{\natexlab{c}}),
  \eprint{1310.4373}.

\bibitem[{\citenamefont{Dong et~al.}(2013{\natexlab{a}})\citenamefont{Dong,
  Faessler, Gutsche, and Lyubovitskij}}]{Dong:2013iqa}
\bibinfo{author}{\bibfnamefont{Y.}~\bibnamefont{Dong}},
  \bibinfo{author}{\bibfnamefont{A.}~\bibnamefont{Faessler}},
  \bibinfo{author}{\bibfnamefont{T.}~\bibnamefont{Gutsche}}, \bibnamefont{and}
  \bibinfo{author}{\bibfnamefont{V.~E.} \bibnamefont{Lyubovitskij}},
  \bibinfo{journal}{Phys. Rev.} \textbf{\bibinfo{volume}{D88}},
  \bibinfo{pages}{014030} (\bibinfo{year}{2013}{\natexlab{a}}),
  \eprint{1306.0824}.

\bibitem[{\citenamefont{Dong et~al.}(2013{\natexlab{b}})\citenamefont{Dong,
  Faessler, Gutsche, and Lyubovitskij}}]{Dong:2013rsa}
\bibinfo{author}{\bibfnamefont{Y.}~\bibnamefont{Dong}},
  \bibinfo{author}{\bibfnamefont{A.}~\bibnamefont{Faessler}},
  \bibinfo{author}{\bibfnamefont{T.}~\bibnamefont{Gutsche}}, \bibnamefont{and}
  \bibinfo{author}{\bibfnamefont{V.~E.} \bibnamefont{Lyubovitskij}},
  \bibinfo{journal}{Few Body Syst.} \textbf{\bibinfo{volume}{54}},
  \bibinfo{pages}{1011} (\bibinfo{year}{2013}{\natexlab{b}}).

\bibitem[{\citenamefont{Dong et~al.}(2013{\natexlab{c}})\citenamefont{Dong,
  Faessler, Gutsche, and Lyubovitskij}}]{Dong:2012hc}
\bibinfo{author}{\bibfnamefont{Y.}~\bibnamefont{Dong}},
  \bibinfo{author}{\bibfnamefont{A.}~\bibnamefont{Faessler}},
  \bibinfo{author}{\bibfnamefont{T.}~\bibnamefont{Gutsche}}, \bibnamefont{and}
  \bibinfo{author}{\bibfnamefont{V.~E.} \bibnamefont{Lyubovitskij}},
  \bibinfo{journal}{J. Phys.} \textbf{\bibinfo{volume}{G40}},
  \bibinfo{pages}{015002} (\bibinfo{year}{2013}{\natexlab{c}}),
  \eprint{1203.1894}.

\bibitem[{\citenamefont{Dong et~al.}(2011{\natexlab{b}})\citenamefont{Dong,
  Faessler, Gutsche, Kumano, and Lyubovitskij}}]{Dong:2011ys}
\bibinfo{author}{\bibfnamefont{Y.}~\bibnamefont{Dong}},
  \bibinfo{author}{\bibfnamefont{A.}~\bibnamefont{Faessler}},
  \bibinfo{author}{\bibfnamefont{T.}~\bibnamefont{Gutsche}},
  \bibinfo{author}{\bibfnamefont{S.}~\bibnamefont{Kumano}}, \bibnamefont{and}
  \bibinfo{author}{\bibfnamefont{V.~E.} \bibnamefont{Lyubovitskij}},
  \bibinfo{journal}{Phys. Rev.} \textbf{\bibinfo{volume}{D83}},
  \bibinfo{pages}{094005} (\bibinfo{year}{2011}{\natexlab{b}}),
  \eprint{1103.4762}.

\bibitem[{\citenamefont{Dong et~al.}(2010{\natexlab{a}})\citenamefont{Dong,
  Faessler, Gutsche, Kumano, and Lyubovitskij}}]{Dong:2010xv}
\bibinfo{author}{\bibfnamefont{Y.}~\bibnamefont{Dong}},
  \bibinfo{author}{\bibfnamefont{A.}~\bibnamefont{Faessler}},
  \bibinfo{author}{\bibfnamefont{T.}~\bibnamefont{Gutsche}},
  \bibinfo{author}{\bibfnamefont{S.}~\bibnamefont{Kumano}}, \bibnamefont{and}
  \bibinfo{author}{\bibfnamefont{V.~E.} \bibnamefont{Lyubovitskij}},
  \bibinfo{journal}{Phys. Rev.} \textbf{\bibinfo{volume}{D82}},
  \bibinfo{pages}{034035} (\bibinfo{year}{2010}{\natexlab{a}}),
  \eprint{1006.4018}.

\bibitem[{\citenamefont{Dong et~al.}(2010{\natexlab{b}})\citenamefont{Dong,
  Faessler, Gutsche, and Lyubovitskij}}]{Dong:2010gu}
\bibinfo{author}{\bibfnamefont{Y.}~\bibnamefont{Dong}},
  \bibinfo{author}{\bibfnamefont{A.}~\bibnamefont{Faessler}},
  \bibinfo{author}{\bibfnamefont{T.}~\bibnamefont{Gutsche}}, \bibnamefont{and}
  \bibinfo{author}{\bibfnamefont{V.~E.} \bibnamefont{Lyubovitskij}},
  \bibinfo{journal}{Phys. Rev.} \textbf{\bibinfo{volume}{D81}},
  \bibinfo{pages}{074011} (\bibinfo{year}{2010}{\natexlab{b}}),
  \eprint{1002.0218}.

\bibitem[{\citenamefont{Dong et~al.}(2010{\natexlab{c}})\citenamefont{Dong,
  Faessler, Gutsche, and Lyubovitskij}}]{Dong:2009tg}
\bibinfo{author}{\bibfnamefont{Y.}~\bibnamefont{Dong}},
  \bibinfo{author}{\bibfnamefont{A.}~\bibnamefont{Faessler}},
  \bibinfo{author}{\bibfnamefont{T.}~\bibnamefont{Gutsche}}, \bibnamefont{and}
  \bibinfo{author}{\bibfnamefont{V.~E.} \bibnamefont{Lyubovitskij}},
  \bibinfo{journal}{Phys. Rev.} \textbf{\bibinfo{volume}{D81}},
  \bibinfo{pages}{014006} (\bibinfo{year}{2010}{\natexlab{c}}),
  \eprint{0910.1204}.

\bibitem[{\citenamefont{Dong et~al.}(2017{\natexlab{b}})\citenamefont{Dong,
  Faessler, and Lyubovitskij}}]{Dong:2017gaw}
\bibinfo{author}{\bibfnamefont{Y.}~\bibnamefont{Dong}},
  \bibinfo{author}{\bibfnamefont{A.}~\bibnamefont{Faessler}}, \bibnamefont{and}
  \bibinfo{author}{\bibfnamefont{V.~E.} \bibnamefont{Lyubovitskij}},
  \bibinfo{journal}{Prog. Part. Nucl. Phys.} \textbf{\bibinfo{volume}{94}},
  \bibinfo{pages}{282} (\bibinfo{year}{2017}{\natexlab{b}}).

\bibitem[{\citenamefont{Tanabashi et~al.}(2018)}]{Tanabashi:2018oca}
\bibinfo{author}{\bibfnamefont{M.}~\bibnamefont{Tanabashi}}
  \bibnamefont{et~al.} (\bibinfo{collaboration}{Particle Data Group}),
  \bibinfo{journal}{Phys. Rev.} \textbf{\bibinfo{volume}{D98}},
  \bibinfo{pages}{030001} (\bibinfo{year}{2018}).

\bibitem[{\citenamefont{Gamermann and Oset}(2007)}]{Gamermann:2007fi}
\bibinfo{author}{\bibfnamefont{D.}~\bibnamefont{Gamermann}} \bibnamefont{and}
  \bibinfo{author}{\bibfnamefont{E.}~\bibnamefont{Oset}},
  \bibinfo{journal}{Eur.Phys.J.} \textbf{\bibinfo{volume}{A33}},
  \bibinfo{pages}{119} (\bibinfo{year}{2007}), \eprint{0704.2314}.

\bibitem[{\citenamefont{Francis et~al.}(2017)\citenamefont{Francis, Hudspith,
  Lewis, and Maltman}}]{Francis:2016hui}
\bibinfo{author}{\bibfnamefont{A.}~\bibnamefont{Francis}},
  \bibinfo{author}{\bibfnamefont{R.~J.} \bibnamefont{Hudspith}},
  \bibinfo{author}{\bibfnamefont{R.}~\bibnamefont{Lewis}}, \bibnamefont{and}
  \bibinfo{author}{\bibfnamefont{K.}~\bibnamefont{Maltman}},
  \bibinfo{journal}{Phys. Rev. Lett.} \textbf{\bibinfo{volume}{118}},
  \bibinfo{pages}{142001} (\bibinfo{year}{2017}), \eprint{1607.05214}.

\bibitem[{\citenamefont{Leskovec et~al.}(2019)\citenamefont{Leskovec, Meinel,
  Pflaumer, and Wagner}}]{Leskovec:2019ioa}
\bibinfo{author}{\bibfnamefont{L.}~\bibnamefont{Leskovec}},
  \bibinfo{author}{\bibfnamefont{S.}~\bibnamefont{Meinel}},
  \bibinfo{author}{\bibfnamefont{M.}~\bibnamefont{Pflaumer}}, \bibnamefont{and}
  \bibinfo{author}{\bibfnamefont{M.}~\bibnamefont{Wagner}},
  \bibinfo{journal}{Phys. Rev.} \textbf{\bibinfo{volume}{D100}},
  \bibinfo{pages}{014503} (\bibinfo{year}{2019}), \eprint{1904.04197}.

\bibitem[{\citenamefont{H{\"o}rz and Hanlon}(2019)}]{Horz:2019rrn}
\bibinfo{author}{\bibfnamefont{B.}~\bibnamefont{H{\"o}rz}} \bibnamefont{and}
  \bibinfo{author}{\bibfnamefont{A.}~\bibnamefont{Hanlon}}
  (\bibinfo{year}{2019}), \eprint{1905.04277}.

\end{thebibliography}
\end{document}